\documentclass[12pt]{article}

\begin{document}

\title{Why the Parity Violation}
\author{G. Quznetsov \\
quznets@yahoo.com}
\date{June 29, 1999}
\maketitle

\begin{abstract}
The sufficient and necessary conditions for a nonzero fermion mass without
Higgs are considered. The Parity Violation is deduced from these conditions.
\end{abstract}

In this paper I consider a global gauge transformation.

I use the following notation:

\[
1_2=\left[ 
\begin{array}{cc}
1 & 0 \\ 
0 & 1
\end{array}
\right] \mbox{,}0_2=\left[ 
\begin{array}{cc}
0 & 0 \\ 
0 & 0
\end{array}
\right] \mbox{, }p_u=\left[ 
\begin{array}{cc}
1 & 0 \\ 
0 & 0
\end{array}
\right] \mbox{, }p_d=\left[ 
\begin{array}{cc}
0 & 0 \\ 
0 & 1
\end{array}
\right] 
\]

\[
\sigma _x=\left[ 
\begin{array}{cc}
0 & 1 \\ 
1 & 0
\end{array}
\right] \mbox{, }\sigma _y=\left[ 
\begin{array}{cc}
0 & -i \\ 
i & 0
\end{array}
\right] \mbox{, }\sigma _z=\left[ 
\begin{array}{cc}
1 & 0 \\ 
0 & -1
\end{array}
\right] , 
\]

\[
\beta _1=\left[ 
\begin{array}{cc}
\sigma _x & 0_2 \\ 
0_2 & -\sigma _x
\end{array}
\right] \mbox{, }\beta _2=\left[ 
\begin{array}{cc}
\sigma _y & 0_2 \\ 
0_2 & -\sigma _y
\end{array}
\right] ,\beta _3=\left[ 
\begin{array}{cc}
\sigma _z & 0_2 \\ 
0_2 & -\sigma _z
\end{array}
\right] , 
\]

\[
\beta _4=i\cdot \left[ 
\begin{array}{cc}
0_2 & 1_2 \\ 
-1_2 & 0_2
\end{array}
\right] \mbox{, }\beta _0=\left[ 
\begin{array}{cc}
1_2 & 0_2 \\ 
0_2 & 1_2
\end{array}
\right] =1_4\mbox{, }\gamma _0=\left[ 
\begin{array}{cc}
0_2 & 1_2 \\ 
1_2 & 0_2
\end{array}
\right] \mbox{,} 
\]

\[
\gamma _5=\left[ 
\begin{array}{cc}
1_2 & 0_2 \\ 
0_2 & -1_2
\end{array}
\right] \mbox{,} 
\]

\[
1_8=\left[ 
\begin{array}{cc}
\beta _0 & 0_4 \\ 
0_4 & \beta _0
\end{array}
\right] \mbox{,} 
\]

\[
0_8=1_8-1_8\mbox{,} 
\]

\[
\gamma _1=\gamma _0\cdot \beta _1\mbox{, }\gamma _2=\gamma _0\cdot \beta _2%
\mbox{, }\gamma _3=\gamma _0\cdot \beta _3\mbox{, }\gamma _5=i\cdot \gamma
_0\cdot \beta _4\mbox{,} 
\]

\[
0_4=\left[ 
\begin{array}{cc}
0_2 & 0_2 \\ 
0_2 & 0_2
\end{array}
\right] \mbox{, }\underline{\gamma _0}=\left[ 
\begin{array}{cc}
\gamma _0 & 0_4 \\ 
0_4 & \gamma _0
\end{array}
\right] \mbox{, }\underline{\beta _n}=\left[ 
\begin{array}{cc}
\beta _n & 0_4 \\ 
0_4 & \beta _n
\end{array}
\right] \mbox{.} 
\]

\section{Hints}

1) Let us consider the free lepton Lagrangian \cite{Rd}:

\[
\mathcal{L}=0.5\cdot i\cdot \left( \overline{\psi }\cdot \gamma ^\mu \cdot
\left( \partial _\mu \psi \right) -\left( \partial _\mu \overline{\psi }%
\right) \cdot \gamma ^\mu \cdot \psi \right) -m\cdot \overline{\psi }\cdot
\psi \mbox{.} 
\]

Hence:

\[
\mathcal{L}=0.5\cdot i\cdot \left( \psi ^{\dagger }\cdot \beta ^\mu \cdot
\left( \partial _\mu \psi \right) -\left( \partial _\mu \psi ^{\dagger
}\right) \cdot \beta ^\mu \cdot \psi \right) -m\cdot \psi ^{\dagger }\cdot
\gamma ^0\cdot \psi \mbox{.} 
\]

This Lagrangian contains four matrices from the Clifford pentad \cite{Md} 
\[
\left\{ \gamma _0\mbox{, }\beta _1\mbox{, }\beta _2\mbox{, }\beta _3\mbox{, }%
\beta _4\right\} \mbox{,} 
\]

but one does not contain $\beta _4$.

2) Let us consider the lepton current:

\[
j_\mu =\psi ^{\dagger }\cdot \beta ^\mu \cdot \psi \mbox{.} 
\]

for $0\leq \mu \leq 3$.

Let us denote:

\[
J_\gamma =\psi ^{\dagger }\cdot \gamma ^0\cdot \psi \mbox{ and }J_4=\psi
^{\dagger }\cdot \beta ^4\cdot \psi \mbox{.} 
\]

In this case if

\[
\rho =j_0 
\]

then the average velocity vector is:

\[
\rho \cdot v_x=j_1\mbox{, }\rho \cdot v_y=j_2\mbox{, }\rho \cdot v_z=j_3%
\mbox{.} 
\]

Let us denote:

\[
\rho \cdot V_\gamma =J_\gamma \mbox{ and }\rho \cdot V_4=J_4\mbox{.} 
\]

In this case:

\[
v_x^2+v_y^2+v_z^2+V_\gamma ^2+V_4^2=1\mbox{.} 
\]

Hence of only all five elements of the Clifford pentad lends the entire kit
of the velocity components.

3) In the {Standard Model} we have got the following entities:

the right electron field vector $e_R$,

the left electron field vector $e_L$,

the electron field vector $e$ ($e=\left[ 
\begin{array}{c}
e_L \\ 
e_R
\end{array}
\right] $),

the left neutrino fields vector $\nu _L$.

the zero right neutrino fields vector $\nu _R$.

the unitary $2\times 2$ $SU(2)$ matrix $U$ of the isospin transformation:

\[
U=\left[ 
\begin{array}{cc}
\cos \left( \varepsilon \right) +i\cdot n_3\cdot \sin \left( \varepsilon
\right) & \left( i\cdot n_1+n_2\right) \cdot \sin \left( \varepsilon \right)
\\ 
\left( i\cdot n_1-n_2\right) \cdot \sin \left( \varepsilon \right) & \cos
\left( \varepsilon \right) -i\cdot n_3\cdot \sin \left( \varepsilon \right)
\end{array}
\right] \mbox{,} 
\]

$\varepsilon $, $n_1$, $n_2$, $n_3$ are real and:

\[
n_1^2+n_2^2+n_3^2=1\mbox{.} 
\]

This matrix acts on the vectors of the kind:$\left[ 
\begin{array}{c}
\nu _L \\ 
e_L
\end{array}
\right] $.

Therefore, in this theory (the (j,0)+(j,0) representation space:  \cite{DVB}%
, \cite{AV}, \cite{Dva}): if

\[
U=\left[ 
\begin{array}{cc}
u_{1,1} & u_{1,2} \\ 
u_{2,1} & u_{2,2}
\end{array}
\right] 
\]

then the matrix

\begin{equation}
\underline{U}=\left[ 
\begin{array}{cccc}
u_{1,1}\cdot 1_2 & 0_2 & u_{1,2}\cdot 1_2 & 0_2 \\ 
0_2 & 1_2 & 0_2 & 0_2 \\ 
u_{2,1}\cdot 1_2 & 0_2 & u_{2,2}\cdot 1_2 & 0_2 \\ 
0_2 & 0_2 & 0_2 & 1_2
\end{array}
\right]  \label{u1}
\end{equation}

operates on the vector

\[
\left[ 
\begin{array}{c}
\nu _L \\ 
\nu _R. \\ 
e_L \\ 
e_R
\end{array}
\right] 
\]

Because $e_R$, $e_L$, $\nu _L$, $\nu _R$ are the two-component vectors then

\[
\left[ 
\begin{array}{c}
\nu _L \\ 
\nu _R \\ 
e_L \\ 
e_R
\end{array}
\right] \mbox{ is }\left[ 
\begin{array}{c}
\nu _{L1} \\ 
\nu _{L2} \\ 
\nu _{R1} \\ 
\nu _{R1} \\ 
e_{L1} \\ 
e_{L2} \\ 
e_{R1} \\ 
e_{R1}
\end{array}
\right] 
\]

$\underline{U}$ has got eight orthogonal normalized eigenvectors $s_1$, $s_2$%
, $s_3$, $s_4$, $s_5$, $s_6$, $s_7$, $s_8$:

\[
s_1=\left[ 
\begin{array}{c}
0 \\ 
0 \\ 
1 \\ 
0 \\ 
0 \\ 
0 \\ 
0 \\ 
0
\end{array}
\right] \mbox{, }s_2=\left[ 
\begin{array}{c}
0 \\ 
0 \\ 
0 \\ 
1 \\ 
0 \\ 
0 \\ 
0 \\ 
0
\end{array}
\right] \mbox{, }s_3=\left[ 
\begin{array}{c}
a \\ 
0 \\ 
0 \\ 
0 \\ 
b+i\cdot c \\ 
0 \\ 
0 \\ 
0
\end{array}
\right] \mbox{, }s_4=\left[ 
\begin{array}{c}
0 \\ 
a \\ 
0 \\ 
0 \\ 
0 \\ 
b+i\cdot c \\ 
0 \\ 
0
\end{array}
\right] \mbox{,} 
\]

\[
s_5=\left[ 
\begin{array}{c}
0 \\ 
0 \\ 
0 \\ 
0 \\ 
0 \\ 
0 \\ 
1 \\ 
0
\end{array}
\right] \mbox{, }s_6=\left[ 
\begin{array}{c}
0 \\ 
0 \\ 
0 \\ 
0 \\ 
0 \\ 
0 \\ 
0 \\ 
1
\end{array}
\right] \mbox{, }s_7=\left[ 
\begin{array}{c}
-b+i\cdot c \\ 
0 \\ 
0 \\ 
0 \\ 
a \\ 
0 \\ 
0 \\ 
0
\end{array}
\right] \mbox{, }s_8=\left[ 
\begin{array}{c}
0 \\ 
-b+i\cdot c \\ 
0 \\ 
0 \\ 
0 \\ 
a \\ 
0 \\ 
0
\end{array}
\right] 
\]

($a$, $b$, $c$ are a real numbers) with the corresponding eigenvalues: $1$, $%
1$, $\exp \left( i\cdot \lambda \right) $, $\exp \left( i\cdot \lambda
\right) $, $1$, $1$, $\exp \left( -i\cdot \lambda \right) $, $\exp \left(
-i\cdot \lambda \right) $.

Let: $K$ be the $8\times 8$ complex matrix, constructed by $s_1$, $s_2$, $%
s_3 $, $s_4$, $s_5$, $s_6$, $s_7$, $s_8$ as the following:

\[
K=\left[ 
\begin{array}{cccccccc}
s_1 & s_2 & s_3 & s_4 & s_5 & s_6 & s_7 & s_8
\end{array}
\right] \mbox{.} 
\]

Let for all $k$ ($1\leq k\leq 8$):

\[
h_k=\underline{\gamma _0}\cdot s_k 
\]

and let:

\[
M=\left[ 
\begin{array}{cccccccc}
h_1 & h_2 & h_3 & h_4 & h_5 & h_6 & h_7 & h_8
\end{array}
\right] \mbox{.} 
\]

Let:

\[
P_3=\left[ 
\begin{array}{cccc}
0_2 & 0_2 & 0_2 & 0_2 \\ 
0_2 & p_u & 0_2 & 0_2 \\ 
0_2 & 0_2 & 0_2 & 0_2 \\ 
0_2 & 0_2 & 0_2 & 0_2
\end{array}
\right] \mbox{,}P_4=\left[ 
\begin{array}{cccc}
0_2 & 0_2 & 0_2 & 0_2 \\ 
0_2 & p_d & 0_2 & 0_2 \\ 
0_2 & 0_2 & 0_2 & 0_2 \\ 
0_2 & 0_2 & 0_2 & 0_2
\end{array}
\right] \mbox{,} 
\]

\[
P_7=\left[ 
\begin{array}{cccc}
0_2 & 0_2 & 0_2 & 0_2 \\ 
0_2 & 0_2 & 0_2 & 0_2 \\ 
0_2 & 0_2 & 0_2 & 0_2 \\ 
0_2 & 0_2 & 0_2 & p_u
\end{array}
\right] \mbox{, }P_8=\left[ 
\begin{array}{cccc}
0_2 & 0_2 & 0_2 & 0_2 \\ 
0_2 & 0_2 & 0_2 & 0_2 \\ 
0_2 & 0_2 & 0_2 & 0_2 \\ 
0_2 & 0_2 & 0_2 & p_d
\end{array}
\right] \mbox{.} 
\]

In this case the projection matrices are:

\[
\begin{array}{c}
Y_1=M\cdot P_3\cdot M^{\dagger }\mbox{,} \\ 
Y_2=M\cdot P_4\cdot M^{\dagger }\mbox{,} \\ 
Y_3=K\cdot P_3\cdot K^{\dagger }\mbox{,} \\ 
Y_4=K\cdot P_4\cdot K^{\dagger }\mbox{,} \\ 
Y_5=M\cdot P_7\cdot M^{\dagger }\mbox{,} \\ 
Y_6=M\cdot P_8\cdot M^{\dagger }\mbox{,} \\ 
Y_7=K\cdot P_7\cdot K^{\dagger }\mbox{,} \\ 
Y_8=K\cdot P_8\cdot K^{\dagger }\mbox{.}
\end{array}
\]

The vectors:

\[
\underline{e}=\left[ 
\begin{array}{c}
0 \\ 
0 \\ 
0 \\ 
0 \\ 
e_{L1} \\ 
e_{L2} \\ 
e_{R1} \\ 
e_{R2}
\end{array}
\right] \mbox{, }\underline{e_R}=\left[ 
\begin{array}{c}
0 \\ 
0 \\ 
0 \\ 
0 \\ 
0 \\ 
0 \\ 
e_{R1} \\ 
e_{R2}
\end{array}
\right] \mbox{, }\underline{e_L}=\left[ 
\begin{array}{c}
0 \\ 
0 \\ 
0 \\ 
0 \\ 
e_{L1} \\ 
e_{L2} \\ 
0 \\ 
0
\end{array}
\right] \mbox{.} 
\]

correspond to the vectors $e$, $e_R$ and $e_L$ resp.

Let:

\[
\begin{array}{c}
X_a=Y_1+Y_2+Y_3+Y_4\mbox{,} \\ 
X_b=Y_5+Y_6+Y_7+Y_8\mbox{,} \\ 
\underline{e_a}=X_a\cdot \underline{e}\mbox{,} \\ 
\underline{e_b}=X_b\cdot \underline{e}\mbox{.}
\end{array}
\]

In this case:

\[
\begin{array}{c}
X_a+X_b=1_8\mbox{,} \\ 
X_a\cdot X_b=0_8\mbox{,} \\ 
X_a\cdot X_a=X_a\mbox{,} \\ 
X_b\cdot X_b=X_b\mbox{,} \\ 
X_a^{\dagger }=X_a\mbox{,} \\ 
X_b^{\dagger }=X_b\mbox{.}
\end{array}
\]

Let:

\[
\begin{array}{c}
\rho _a=\underline{e_a}^{\dagger }\cdot \underline{e_a}\mbox{, }\rho _b=%
\underline{e_b}^{\dagger }\cdot \underline{e_b}\mbox{,} \\ 
J_{\gamma ,a}=\underline{e_a}^{\dagger }\cdot \underline{\gamma _0}\cdot 
\underline{e_a}\mbox{, }J_{\gamma ,b}=\underline{e_b}^{\dagger }\cdot 
\underline{\gamma _0}\cdot \underline{e_b}\mbox{,} \\ 
J_{4,a}=\underline{e_a}^{\dagger }\cdot \underline{\beta _4}\cdot \underline{%
e_a}\mbox{, }J_{4,b}=\underline{e_b}^{\dagger }\cdot \underline{\beta _4}%
\cdot \underline{e_b}\mbox{,} \\ 
J_{\gamma ,a}=\rho _a\cdot V_{\gamma ,a}\mbox{, }J_{\gamma ,b}=\rho _b\cdot
V_{\gamma ,b}\mbox{,} \\ 
J_{4,a}=\rho _a\cdot V_{4,a}\mbox{, }J_{4,b}=\rho _b\cdot V_{4,b}\mbox{.}
\end{array}
\]

Let:

\[
\begin{array}{c}
\underline{e_a}^{\prime }=U\cdot \underline{e_a}\mbox{, }\underline{e_b}%
^{\prime }=U\cdot \underline{e_b} \\ 
\rho _a^{\prime }=\underline{e_a}^{\prime \dagger }\cdot \underline{e_a}%
^{\prime }\mbox{, }\rho _b^{\prime }=\underline{e_b}^{\prime \dagger }\cdot 
\underline{e_b}^{\prime }\mbox{,} \\ 
J_{\gamma ,a}^{\prime }=\underline{e_a}^{\prime \dagger }\cdot \underline{%
\gamma _0}\cdot \underline{e_a}^{\prime }\mbox{, }J_{\gamma ,b}^{\prime }=%
\underline{e_b}^{\prime \dagger }\cdot \underline{\gamma _0}\cdot \underline{%
e_b}^{\prime }\mbox{,} \\ 
J_{4,a}^{\prime }=\underline{e_a}^{\prime \dagger }\cdot \underline{\beta _4}%
\cdot \underline{e_a}^{\prime }\mbox{, }J_{4,b}^{\prime }=\underline{e_b}%
^{\prime \dagger }\cdot \underline{\beta _4}\cdot \underline{e_b}^{\prime }%
\mbox{,} \\ 
J_{\gamma ,a}^{\prime }=\rho _a^{\prime }\cdot V_{\gamma ,a}^{\prime }%
\mbox{, }J_{\gamma ,b}^{\prime }=\rho _b^{\prime }\cdot V_{\gamma
,b}^{\prime }\mbox{,} \\ 
J_{4,a}^{\prime }=\rho _a^{\prime }\cdot V_{4,a}^{\prime }\mbox{, }%
J_{4,b}^{\prime }=\rho _b^{\prime }\cdot V_{4,b}^{\prime }\mbox{.}
\end{array}
\]

In this case:

\[
V_{\gamma ,a}=V_{\gamma ,b}\mbox{, }V_{4,a}=V_{4,b} 
\]

but:

\[
\begin{array}{c}
V_{\gamma ,a}^{\prime }=V_{\gamma ,a}\cdot \cos \left( \lambda \right)
-V_{4,a}\cdot \sin \left( \lambda \right) \mbox{,} \\ 
V_{4,a}^{\prime }=V_{4,a}\cdot \cos \left( \lambda \right) +V_{\gamma
,a}\cdot \sin \left( \lambda \right) \mbox{,} \\ 
V_{\gamma ,b}^{\prime }=V_{\gamma ,b}\cdot \cos \left( \lambda \right)
+V_{4,b}\cdot \sin \left( \lambda \right) \mbox{,} \\ 
V_{4,b}^{\prime }=V_{4,b}\cdot \cos \left( \lambda \right) -V_{\gamma
,b}\cdot \sin \left( \lambda \right) \mbox{.}
\end{array}
\]

Hence, every isospin transformation $U$ divides a electron on two components
which scatter on the angle $2\cdot \lambda $ in the space of ($J_\gamma $, $%
J_4$).

Hence $\beta ^4$ must be inserted into Lagrangian.

\section{Sufficient Conditions}

Let $\underline{\psi }$ be any field of the following type:

\[
\underline{\psi }=\left[ 
\begin{array}{c}
0 \\ 
0 \\ 
0 \\ 
0 \\ 
\psi _{L1} \\ 
\psi _{L2} \\ 
\psi _{R1} \\ 
\psi _{R2}
\end{array}
\right] \mbox{.} 
\]

The value of the form

\begin{equation}
\begin{array}{c}
\left( \left( \underline{\psi }^{\dagger }\cdot \underline{\gamma ^0}\cdot
X_a\cdot \underline{\psi }\right) ^2+\left( \underline{\psi }^{\dagger
}\cdot \underline{\beta ^4}\cdot X_a\cdot \underline{\psi }\right) ^2\right)
^{0.5}+ \\ 
+\left( \left( \underline{\psi }^{\dagger }\cdot \gamma ^0\cdot X_b\cdot 
\underline{\psi }\right) ^2+\left( \underline{\psi }^{\dagger }\cdot \beta
^4\cdot X_b\cdot \underline{\psi }\right) ^2\right) ^{0.5}
\end{array}
\label{u2}
\end{equation}

does not depend from the choice of the $SU(2)$ matrix $U$ and the Lagrangian:

\[
\begin{array}{c}
\mathcal{L}_\psi =0.5\cdot i\cdot \left( \underline{\psi }^{\dagger }\cdot 
\underline{\beta ^\mu }\cdot \left( \partial _\mu \underline{\psi }\right)
-\left( \partial _\mu \underline{\psi }\right) ^{\dagger }\cdot \underline{%
\beta ^\mu }\cdot \underline{\psi }\right) -\  \\ 
-m_\psi \cdot \left( 
\begin{array}{c}
\left( \left( \underline{\psi }^{\dagger }\cdot \underline{\gamma ^0}\cdot
X_a\cdot \underline{\psi }\right) ^2+\left( \underline{\psi }^{\dagger
}\cdot \underline{\beta ^4}\cdot X_a\cdot \underline{\psi }\right) ^2\right)
^{0.5}+ \\ 
+\left( \left( \underline{\psi }^{\dagger }\cdot \underline{\gamma ^0}\cdot
X_b\cdot \underline{\psi }\right) ^2+\left( \underline{\psi }^{\dagger
}\cdot \underline{\beta ^4}\cdot X_b\cdot \underline{\psi }\right) ^2\right)
^{0.5}
\end{array}
\right)
\end{array}
\]

is invariant for this $SU(2)$ transformation.

Let us denote:

\begin{eqnarray*}
\frac{\left( \underline{\psi }^{\dagger }\cdot \underline{\gamma ^0}\cdot
X_a\cdot \underline{\psi }\right) }{\sqrt{\left( \underline{\psi }^{\dagger
}\cdot \underline{\gamma ^0}\cdot X_a\cdot \underline{\psi }\right)
^2+\left( \underline{\psi }^{\dagger }\cdot \underline{\beta ^4}\cdot
X_a\cdot \underline{\psi }\right) ^2}} &=&\cos \left( \alpha _a\right) %
\mbox{,} \\
\frac{\left( \underline{\psi }^{\dagger }\cdot \underline{\beta ^4}\cdot
X_a\cdot \underline{\psi }\right) }{\sqrt{\left( \underline{\psi }^{\dagger
}\cdot \underline{\gamma ^0}\cdot X_a\cdot \underline{\psi }\right)
^2+\left( \underline{\psi }^{\dagger }\cdot \underline{\beta ^4}\cdot
X_a\cdot \underline{\psi }\right) ^2}} &=&\sin \left( \alpha _a\right) %
\mbox{,}
\end{eqnarray*}

\begin{eqnarray*}
\frac{\left( \underline{\psi }^{\dagger }\cdot \underline{\gamma ^0}\cdot
X_b\cdot \underline{\psi }\right) }{\sqrt{\left( \underline{\psi }^{\dagger
}\cdot \underline{\gamma ^0}\cdot X_b\cdot \underline{\psi }\right)
^2+\left( \underline{\psi }^{\dagger }\cdot \underline{\beta ^4}\cdot
X_b\cdot \underline{\psi }\right) ^2}} &=&\cos \left( \alpha _b\right) %
\mbox{.} \\
\frac{\left( \underline{\psi }^{\dagger }\cdot \underline{\beta ^4}\cdot
X_b\cdot \underline{\psi }\right) }{\sqrt{\left( \underline{\psi }^{\dagger
}\cdot \underline{\gamma ^0}\cdot X_b\cdot \underline{\psi }\right)
^2+\left( \underline{\psi }^{\dagger }\cdot \underline{\beta ^4}\cdot
X_b\cdot \underline{\psi }\right) ^2}} &=&\sin \left( \alpha _b\right)
\end{eqnarray*}

Let:

\[
\underline{\gamma }=\left( \cos \left( \alpha _a\right) \cdot \underline{%
\gamma ^0}+\sin \left( \alpha _a\right) \cdot \underline{\beta ^4}\right)
\cdot X_a+\left( \cos \left( \alpha _b\right) \cdot \underline{\gamma ^0}%
+\sin \left( \alpha _b\right) \cdot \underline{\beta ^4}\right) \cdot X_b%
\mbox{.} 
\]

In this case:

\[
\underline{\gamma }\cdot \underline{\gamma }=1_8 
\]

and if $1\leq k\leq 3$ then

\[
\underline{\gamma }\cdot \underline{\beta ^k}=-\underline{\beta ^k}\cdot 
\underline{\gamma } 
\]

and the Euler-Lagrange equation for $\mathcal{L}_\psi $ is the following:

\[
\left( i\cdot \underline{\beta ^\mu }\cdot \partial _\mu -m_\psi \underline{%
\cdot \gamma }\right) \cdot \underline{\psi }=0\mbox{.} 
\]

Since

\[
\alpha _a=\alpha _b 
\]

then

\[
\underline{\gamma }=\cos \left( \alpha _a\right) \cdot \underline{\gamma ^0}%
+\sin \left( \alpha _a\right) \cdot \underline{\beta ^4}\mbox{.} 
\]

Let $\underline{\psi }$ be a plane wave electron spinor \cite{Rd} with a
positive energy:

\[
\underline{\psi }=\left( a_1\cdot \left[ 
\begin{array}{c}
0 \\ 
0 \\ 
0 \\ 
0 \\ 
1 \\ 
0 \\ 
\frac{p_z}{E+m_e} \\ 
\frac{p_x+i\cdot p_y}{E+m_e}
\end{array}
\right] +a_2\cdot \left[ 
\begin{array}{c}
0 \\ 
0 \\ 
0 \\ 
0 \\ 
0 \\ 
1 \\ 
\frac{p_x-i\cdot p_y}{E+m_e} \\ 
\frac{-p_z}{E+m_e}
\end{array}
\right] \right) \cdot \exp \left( -i\cdot p\cdot x\right) \mbox{,} 
\]

here:

$a_1$, $a_2$ are complex, and $E=\sqrt{p^2+m_e^2}$.

In this case:

\[
\cos \left( \alpha _a\right) =1\mbox{.} 
\]

Hence

\[
\underline{\gamma }=\underline{\gamma ^0} 
\]

and the Euler-Lagrange equation is the following:

\[
\left( i\cdot \underline{\gamma ^\mu }\cdot \partial _\mu -m_\psi \right)
\cdot \underline{\psi }=0\mbox{.}. 
\]

\section{Necessary Conditions}

Let \underline{$U$} be any $8\times 8$ complex matrix for which the
Lagrangian

\[
\mathcal{L}_0=0.5\cdot i\cdot \left( \underline{\psi }^{\dagger }\cdot 
\underline{\beta ^\mu }\cdot \left( \partial _\mu \underline{\psi }\right)
-\left( \partial _\mu \underline{\psi }\right) ^{\dagger }\cdot \underline{%
\beta ^\mu }\cdot \underline{\psi }\right) 
\]

is invariant. Hence

\[
\underline{U}^{\dagger }\cdot \underline{U}=1_8 
\]

and for all $\mu $ ($1\leq \mu \leq 3$):

\[
\underline{U}\cdot \underline{\beta ^\mu }=\underline{\beta ^\mu }\cdot 
\underline{U}\mbox{.} 
\]

$\underline{U}$ must be of the following type from this commutativity :

\[
\underline{U}=\left[ 
\begin{array}{cccccccc}
z_{1,1} & 0 & 0 & 0 & z_{1,5} & 0 & 0 & 0 \\ 
0 & z_{1,1} & 0 & 0 & 0 & z_{1,5} & 0 & 0 \\ 
0 & 0 & z_{3,3} & 0 & 0 & 0 & z_{3,7} & 0 \\ 
0 & 0 & 0 & z_{3,3} & 0 & 0 & 0 & z_{3,7} \\ 
z_{5,1} & 0 & 0 & 0 & z_{5,5} & 0 & 0 & 0 \\ 
0 & z_{5,1} & 0 & 0 & 0 & z_{5,5} & 0 & 0 \\ 
0 & 0 & z_{7,3} & 0 & 0 & 0 & z_{7,7} & 0 \\ 
0 & 0 & 0 & z_{7,3} & 0 & 0 & 0 & z_{7,7}
\end{array}
\right] 
\]

(here $z_{j,k}$ are a complex) and from the unitarity if

\[
z_{j,k}=x_{j,k}+i\cdot y_{j,k} 
\]

then

\[
\begin{array}{c}
1-x_{1,5}^2-y_{1,5}^2-y_{5,5}^2\geq 0\mbox{,} \\ 
x_{1,1}=\sqrt{1-x_{1,5}^2-y_{1,5}^2-y_{5,5}^2}\mbox{,} \\ 
x_{5,5}=x_{1,1}\mbox{,} \\ 
x_{5,1}=-x_{1,5}\mbox{,} \\ 
y_{1,1}=-y_{5,5}\mbox{,} \\ 
y_{5,1}=y_{1,5}
\end{array}
\]

and

\[
\begin{array}{c}
1-x_{3,7}^2-y_{3,7}^2-y_{7,7}^2\geq 0\mbox{,} \\ 
x_{3,3}=\sqrt{1-x_{3,7}^2-y_{3,7}^2-y_{7,7}^2}\mbox{,} \\ 
x_{7,7}=x_{3,3}\mbox{,} \\ 
x_{7,3}=-x_{3,7}\mbox{,} \\ 
y_{3,3}=-y_{7,7}\mbox{,} \\ 
y_{7,3}=y_{3,7}\mbox{.}
\end{array}
\]

If

\[
\begin{array}{c}
x_{3,7}=0\mbox{,} \\ 
y_{3,7}=0\mbox{,} \\ 
y_{7,7}=0
\end{array}
\]

then

\[
z_{3,3}=1 
\]

and $\underline{U}$ is the matrix of type (\ref{u1}). In this case the mass
form (\ref{u2}) is invariant for $\underline{U}$ and a right-handed
particles do not interact by this transformation.

Therefore if an electron has got a nonzero mass, provided with the mass form
(\ref{u2}), then all neutrinos must be left-handed.

Like this, for $z_{1,1}=1$, all antineutrinos must be right-handed to an
positron has got a nonzero mass.

If $z_{1,1}\neq 1$ and $z_{3,3}\neq 1$ then a mass form, invariant for $%
\underline{U}$, does not exist.

\section{{\bf Acknowledgment}}

Thanks very much to Prof. V. V. Dvoeglazov for his papers which he had sent
me kindly in 1998.

\end{document}